\documentstyle[11pt,newpasp,twoside,epsf]{article}
\begin{document}
\title{Stellar Abundances in Local Group Galaxies with the VLT}
 \author{Eline Tolstoy}
\affil{Kapteyn Institute, Univ. of Groningen, the Netherlands}
\author{Kim Venn}
\affil{Macalester College \& Univ. of Minnesota, USA}

\begin{abstract}
Here we describe some of our latest results from measuring detailed
abundances in Local Group dwarf galaxies with VLT.  Combining
spectroscopic abundances with Colour-Magnitude diagrams allows the
effective {\it measurement} of detailed chemical evolution with time
in these galaxies. Although there are not yet significant numbers of
individual stars observed in local group dwarf galaxies the uniformity
of the abundance patterns of the majority of stars in galaxies with
very different star formation histories must hint at a general
properties of all star formation in these small systems.
\end{abstract}

\section{The Fossil Record of the Early Universe}

Low mass stars, such as those found on the Red Giant Branch (RGB),
have long lifetimes, often comparable to the age of the Universe, and
they have retained much of their original chemical composition in
their atmospheres.  These stars are thus very useful because they are
fossils containing a direct measure of abundances and their evolution
since the earliest times . This is what Freeman \& Bland-Hawthorn
(2002) have neatly termed {\it Chemical Tagging}.

The spectroscopic abundance measurements of individual stars on the
RGB by definition break the age-metallicity degeneracy in determining
the star formation histories of galaxies from Colour-Magnitude
Diagrams (CMDs). High resolution spectra give a much more accurate
measure of the metallicity than can be obtained from CMD analysis
alone. If we believe that theoretical stellar evolution tracks are
reasonably accurate, then we can determine ages for all the stars for
which we have measured a metallicity. These ages range from the
earliest star formation in the Universe to 1$-$2~Gyr from the present
day. Thus they enable a measurement of the chemical evolution
variations within dwarf galaxies over almost the entire age of the
(star-forming) Universe.  With a large enough sample of stars we can
determine the influence of gas and metal infall and outflow in the
galaxy.  We can determine the relative importance of different
enrichment processes over time (e.g., Supernovae Ia and II, AGB stars,
stellar winds and the like).  We can re-examine our understanding
of the nucleosynthesis of the elements, especially those that have
many possible formation sites, {\it e.g.,} Ti, Mg/Ca, Cu/Fe, Zn/Fe,
Y/Ba, Ba/Eu, Mg/Eu.

It is also possible to measure accurate abundances of relatively young
stars in nearby galaxies to determine how the ongoing star formation
we observe directly today in dwarf galaxies relates to past star
formation. Dwarf galaxies can have extremely low metallicities and
therefore unable us to study star formation in a regime which is more
common at high-redshift.

As well as allowing us a better understanding of star formation and
the effects it has on individual systems spectroscopic abundances
allow a comparison between the chemical signatures of stars found in
dwarf galaxies with stars found in other environments, such as our
Galaxy.  This allows us to examine the theory that larger galaxies
like our own were built up from small, dwarf galaxy sized clumps.  It
is clear from direct observations of the Universe at all redshifts
that galaxies frequently merge with each other. Looking at the
differences in properties of stars in small and large galaxies
provides additional constraints on the likelihood and time scale of
these merger events.

Dwarf galaxy size objects are believed to be the first structures to
form in the early universe and thus they are potentially the sites of
the formation of the first stars. Of course we have no guarantee
that these first structures {\it were} actually dwarf galaxies that
have survived until today but there is no evidence against this
either.  All dwarf galaxies for which sufficiently detailed
observations exist have ancient stellar populations, which are most
directly observed by the presence of a blue Horizontal branch in their
CMD and/or an RR~Lyr variable star population.  There are a number of
scenarios as to how the first stars may have
formed, each with a distinct
chemical signature for which we can test.

\section{Measuring Chemical Evolution}

There are several different approaches to measuring stellar abundances
in nearby galaxies. Each have their own regime of validity and
interest, and all contain the inherent assumption that the stellar
abundances we observe are representative of the abundance of the gas
out of which the stars were initially formed.

For the nearest by dwarf galaxies, out to about 200~kpc from us, we
can obtain high resolution spectra of individual RGB stars. This
provides the most direct information about the chemical evolution of
the host galaxy over the longest time baseline.  These type of
observations have been made with UVES (e.g., Shetrone et al. 2003;
Tolstoy et al. 2003; Hill et al. 2000) and also at Keck with HIRES
(e.g., Shetrone et al. 2001). High resolution spectra allow the
determination of the abundance of a wealth of different chemical
elements often from more than one line.  In each UVES spectrum, for
example, there around 100 lines of iron (Fe~I and Fe~II), there are
also lines from $\alpha$-elements (e.g., O, Ca, Mg, Ti), Fe-peak
elements (V, Cr, Mn), and neutron capture elements (e.g., Y, Ba, La,
Eu). Several of these elements are also frequently observed in high
resolution absorption spectra of high-redshift Damped Lyman-Alpha
(DLA) systems (e.g., Zn, Cr, Mn). Direct comparison can also be made
with extensive spectroscopic surveys of stars in the Milky Way (e.g.,
Edvardsson et al. 1993; McWilliam et al. 1995; Fulbright 2002).

It is also possible to take lower resolution spectra of more simple 
metallicity indicators. For example the Ca~II triplet at
$\lambda\lambda$8500, 8544, 8665~\AA is a well calibrated
[Fe/H] indicator (e.g., Cole et al. 2000, 2003) which allows us to
trace [Fe/H] with time using RGB stars.  This has
been successfully applied to several nearby galaxies, and allows us to
survey galaxies out 1.5~Mpc, or almost the entire volume of the Local
Group (e.g., M31: Reitzel \& Guhathakurta 2002; LMC: 
Olszewski et al. 1991, Cole et
al. 2000; Fornax, Sculptor \& NGC~6822: Tolstoy
et al. 2001).  Although this abundance measurement is quite basic
(only providing information on Fe) many measurements can efficiently
be made and this method 
can benefit greatly from
multi-object instruments (e.g., VIMOS and FORS on VLT and DEMOS on
Keck). In fact the wavelength range of the Ca~II triplet is $<$
200\AA, so a narrow band filter can enhance the multiplexing power of
slit mask instruments many fold.

\begin{figure}
\plotone{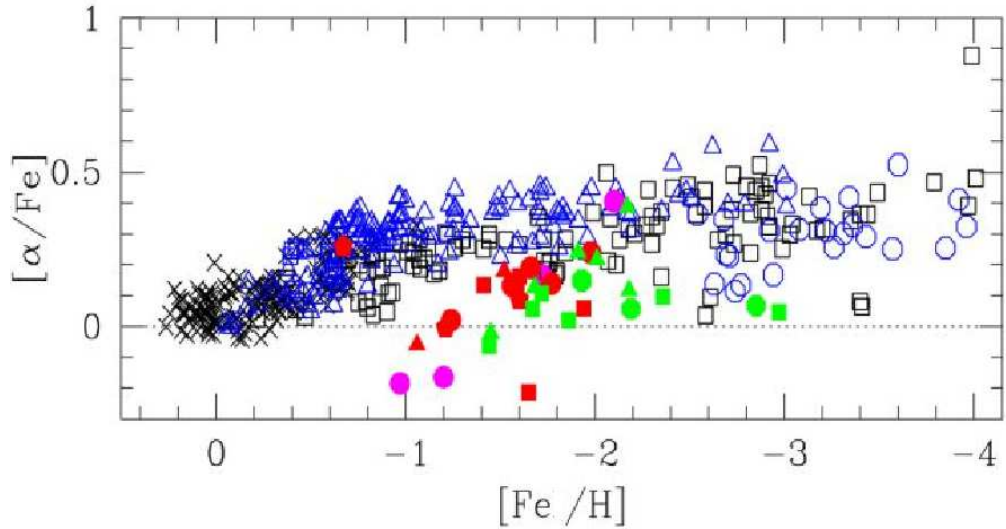}
\caption{$\alpha$-Element abundances.  UVES $\alpha$-abundances from
Tolstoy {\it et al.} 2002 and Shetrone {\it et al.} 2001 plotted
versus [Fe/H] as filled symbols which represent the individual stars
observed in Carina, Leo~I, Sculptor, Fornax, Draco, Ursa Minor and
Sextans dSphs. The crosses are Galactic disk star measurements from
Edvardsson {\it et al.} 1993; the open squares are halo data from
McWilliam {\it et al.} 1995 and the open circles and triangles are
Galactic stars from Ryan et al. (1995) and Fulbright (2002).  This plot highlights the
differences between the $\alpha$-element abundances observed in our
Milky Way and in dwarf galaxies.}
\end{figure}

\begin{figure}
\plotone{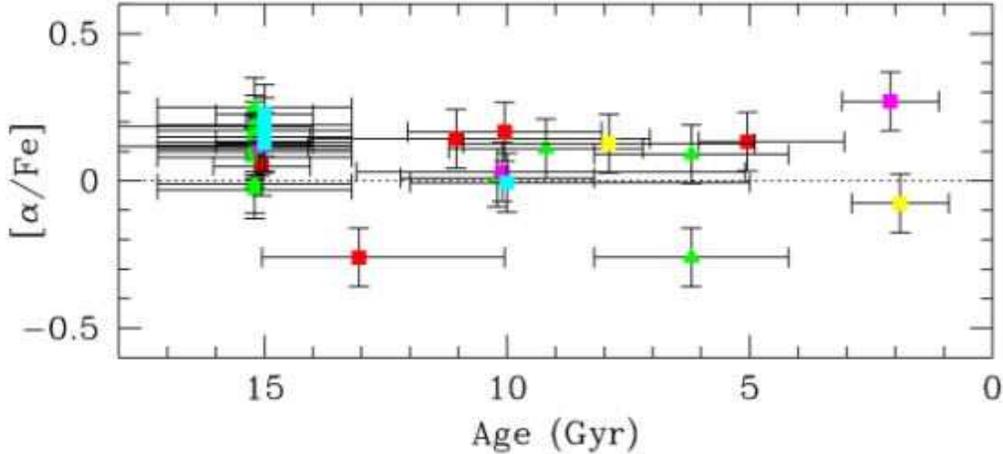}
\caption{$\alpha$-element abundance plotted against time. The age of
each star is determined from an isochrone of the measured metallicity
of the best fitting age. The objects are plotted here are stars from
the same galaxies listed in Figure~1. There is no obvious trend with
age.}
\end{figure}

It is also possible to observe Blue Supergiants (BSGs)
at high resolution to
obtain more detailed abundance information for young stars in galaxies
out to 1.5~Mpc distance. BSGs are much brighter than RGB
stars, but they are also much younger. This means that they provide us with
an accurate measurement of the {\it present day} metallicity in the
galaxies where they are observed but they do not provide a {\it
direct} measurement of chemical evolution, although they are of course
the end product of galaxy evolution.  Long observations with UVES at
VLT (or HIRES on Keck) are typically required for these
abundance studies. There have been several galaxies studied to date
(e.g., M31, NGC~6822, WLM: Venn et al. 2000, 2001, 2003; Sextans A:
Kaufer et al. 2003). These measurements can, of course, only be made
in galaxies with recent star formation, i.e. dwarf irregulars rather
than dwarf spheroidals where most of the detailed RGB abundances have
been measured. Dwarf Irregulars are typically somewhat more massive
than dSph and have arguably had an evolution less disturbed by close
proximity to our Galaxy. These abundances can also be compared to H~II
region metallicities, determined from emission line analysis of 
ionised gas. Both measures of current chemical enrichment in
a galaxy are generally in agreement with each other, although some
enigma remain (e.g., Venn et al. 2003).

\section{Interpreting the Abundance Patterns}

One of the, perhaps, most surprising results of the measurement of
abundance ratios in RGB in dwarf spheroidal galaxies is that the
$\alpha$ elements are typically found to be around solar 
and there is no apparent correlation between age and star
formation history and $\alpha$-abundance (see Figures~1~\&~2). 
If we were to
take the Milky Way as a template (e.g., Gilmore \& Wyse 1991), for
their [Fe/H] and especially for the oldest stars in each system we
would expect dwarf spheroidals to have higher [$\alpha$/Fe], than the
average trend which is
observed in Figure~1.

Although the number of stars with high resolution abundances in any
given dwarf spheroidal to date is on average very small (around 5),
the [$\alpha$/Fe] measured is consistently low for all dSph despite
widely varying ages, [Fe/H] and star formation histories. This
suggests that for dSph the Galaxy is probably not the best template to
interpret the observed abundances.  It seems that low
$\alpha$-abundance need not mean stars which have been made from
material which was enriched by SNIa explosion.  It is possible for a
low-$\alpha$ enrichment pattern to exist from very early on in the star
formation history of a galaxy {\it before there has been any time for
SNIa enrichment} (Tolstoy et al. 2003), 
unless we have significantly overestimated the SNIa
time scale at low metallicities (not impossible).  It is also possible
to invoke AGB wind enrichment, as another mechanism to diminish
[$\alpha$/Fe], and account for the over-abundance of s-process
elements (e.g., Ba, Y).  However, the timescale for this to occur is
also thought to be at least a giga-year after the onset of star
formation and so this is would have to occur more rapidly than current
predictions suggest (also not impossible) to fully explain our
observations.

It is also 
possible that the dwarf galaxies we observed may have an {\it
effectively} truncated Initial Mass Function, in the sense of a
lack of very high mass stars.  This is not too hard to envisage as
these 
dwarf galaxies are small systems with extremely low star formation
rates through out their history. They may not typically form very high
mass molecular clouds, and thus the probability that a galaxy will
form many (or even any) high mass stars is statistically low.  This
could explain the abundance patterns seen.  In addition to the low
$\alpha$-element abundances observed, the enhanced abundance of
r-process element Eu, for example, is 
consistent with the scenario of predominantly low mass SNII in a slow
evolving environment.

It is also possible that blow-out has played a significant role in the
chemical evolution of these galaxies going back to the earliest
times. But it would have to be quite 
selective (predominantly expelling $\alpha$-elements for example), 
and consistent over a
range of different galaxy masses (and types).  Blowing up a small
galaxy is ``easy'' in theory, (e.g., Mori, Ferrara \& Madau 2001) but
it is not so easy to find direct evidence of blow-out in a small
galaxy 
with the confidence
that the gas currently seen flowing out will never 
return (e.g., Martin et al. 2002). Determining
if gas and/or metals will leave a galaxy for good is very sensitive to
the structure of the ISM in a galaxy and whether or not it contains
a significant gaseous halo, and how high the star formation rate can
be at any given time.

It is also interesting to note that [$\alpha$/Fe] measured in massive
(young) BSG stars in dIrr galaxies is also low and thus 
also unlike the Milky
Way at the same [Fe/H] (e.g., Venn et al. 2001, 2003; Kaufer et
al. 2003).  Another class of object with low $\alpha$-abundance
measurements are Damped Lyman Alpha systems (e.g., Nissen et
al. 2003, submitted). That is not to say that they are necessarily
dwarf galaxies but it appears that their chemical evolution has been
similar to that of dwarf galaxies in the Local Group.

Looking at Figure~1 it is rather clear that significant numbers of
stars from dwarf galaxies cannot be included in in a merging formation
scenario for our Galaxy {\it at any epoch}.  Thus, dwarf galaxies {\it
are not obvious hierarchical fragments}.  The only way this aspect of
standard hierarchical galaxy formation scenario can be retained is if
the dwarf galaxies merged to form larger objects like the Milky Way
{\it very} early in the Universe, before the majority of their stars
were formed, while they were still gas rich.

\section{In Summary}

The results of our observations of high resolution abundances of a
handful of individual stars in nearby dwarf galaxies suggest that
the evolution of [Fe/H] with time is consistent with a closed box
chemical evolution scenario, although the evidence is not very
definitive with the current small samples.
The role of outflows remains unclear. We note that
despite wide variations in star formation histories and [Fe/H] in
dwarf galaxies the [$\alpha$/Fe] abundances are very similar and
typically Galactic-disk-like solar values for the majority of stars we
observe. The Iron peak peak elements observed are similar to the Galactic
halo, but
occur at {\it higher} [Fe/H].  Dwarf galaxies are thus
not obvious hierarchical
fragments.
Their slow evolution apparently leads to distinctly
different abundance patterns from those seen in the Milky Way.

To put our tentative results on a firmer basis requires spectra of
many more stars in dwarf galaxies.  We have in VLT/FLAMES the ideal
instrument for this kind of study. We have put together a programme
with the Dwarf Abundance Radial-velocity Team (DART,
http::/www.astro.rug.nl/$^\sim$dart) to use FLAMES to make detail
observations at high resolution of abundances of more than a hundred
stars in each of three nearby dwarf spheroidal galaxies (Sculptor,
Fornax and Sextans). This promises dramatic new data sets in the
near future 
to answer
many of the unresolved issues discussed here.

\acknowledgments 
ET gratefully acknowledges salary and travel funds
from a fellowship of the Royal
Netherlands Academy of Arts and Sciences.

\end{document}